\documentclass[aps,prd,showpacs,nofootinbib]{revtex4}
\usepackage{eufrak}
\begin{document}

\title{Linear constraints from generally covariant systems with quadratic
  constraints}
\date{\today}
\author{Merced Montesinos}
\email{merced@fis.cinvestav.mx}
\affiliation{Departamento de F\'{\i}sica,
Centro de Investigaci\'on y de
Estudios Avanzados del I.P.N.,\\
Av. I.P.N. No. 2508, 07000 Ciudad de M\'exico, M\'exico.}

\author{Jos\'e David Vergara}
\email{vergara@nuclecu.unam.mx}
\affiliation{Instituto de Ciencias Nucleares,
Universidad Nacional Aut\'onoma de M\'exico, 70-543, Ciudad de M\'exico,
M\'exico}

\begin{abstract}
How to make compatible both boundary and gauge conditions for generally
covariant theories using the gauge symmetry generated by first class
constraints is studied. This approach employs {\it finite} gauge
transformations in contrast with previous works which use infinitesimal ones.
Two kinds of variational principles are taken into account; the first one
features non-gauge-invariant actions whereas the second includes fully
gauge-invariant actions. Furthermore, it is shown that it is possible to
rewrite fully gauge-invariant actions featuring first class constraints
quadratic in the momenta into first class constraints linear in the momenta
(and homogeneous in some cases) due to the full gauge invariance of their
actions. This shows that the gauge symmetry present in generally covariant
theories having first class constraints quadratic in the momenta is not of a
different kind with respect to the one of theories with first class
constraints linear in the momenta if fully gauge-invariant actions are taken
into account for the former theories. These ideas are implemented for the
parametrized relativistic free particle, parametrized harmonic oscillator, and
the $SL(2,R)$ model.
\end{abstract}

\maketitle
\section{Introduction}
The compatibility of both gauge and boundary conditions for gauge theories is
a key point that must be satisfied to have a well-defined dynamics. It could
happen that the boundary conditions chosen for a certain action were
incompatible with the choice made to fix the gauge freedom of the system. If
this were the case, both gauge and boundary conditions could be made
compatible using the infinitesimal gauge symmetry generated by the first class
constraints \cite{Henneaux1992}. It is important to recall that the gauge
transformation associated with the first class constraints is not a sufficient
condition to achieve that goal; an extra input is needed: the transversality
condition, which precisely allows the use of the gauge transformation in the
method. The infinitesimal procedure is completely systematic, being easily
generalized to field theory \cite{MonVer00}. However, is there any difference
if finite gauge transformations are taken into account? On one hand, they are
really important in a nonperturbative treatment of the full symmetry of gauge
systems which has relevance both classically and quantum mechanically. On the
other hand, the finite gauge transformations include the large gauge
transformations; these transformations are no connected to the identity,
therefore their effects are not observed by the infinitesimal procedure. So,
for example, the application of the infinitesimal procedure to the $SL(2,R)$
model implies that its diffeomorphism constraint (linear and homogeneous in
the momenta) does not contribute to the boundary term. However, the finite
approach developed here shows that it really does contribute to it. Obviously,
this contribution cannot be neglected if a complete analysis of the full gauge
symmetry is required. In this way, the procedure here presented can be useful
in the analysis of gauge-invariant systems with nontrivial topological spaces
or in systems with singular boundary conditions. That is why {\it finite}
gauge transformations are really important.

To implement in the action the finite gauge transformations, two kinds of
variational principles are analyzed: the first one features
non-gauge-invariant actions whereas the ones of the second are fully
gauge-invariant.

The first type of variational principle based on non-gauge-invariant actions
is conceptually the finite extension of the method of Ref.
\cite{Henneaux1992}. Like there, the transversality condition is not necessary
in the case of noncanonical gauges but is needed if algebraic gauge conditions
are taken into account. Even though ``gauge-invariant" actions are used in
Ref. \cite{Henneaux1992}, the meaning of gauge invariance adopted there is not
the standard one. So, strictly speaking, variational principles with fully
gauge-invariant actions were not considered there either in their
infinitesimal or finite versions.

The second type of variational principle introduced here includes fully
gauge-invariant actions, where gauge invariance has the usual meaning. Of
course, the same issues concerning the transversality condition are those of
Ref. \cite{Henneaux1992} for the same reasons that apply in that case. This
second type of variational principle is analyzed in both its finite and
infinitesimal versions. Finally, these two kinds of variational principles are
applied to the parametrized relativistic free particle, parametrized harmonic
oscillator, and the $SL(2,R)$ model with two non-commuting Hamiltonian
constraints introduced in Ref.\cite{Mon99}. This is the first result of the
paper displayed in detail in Sec. II.

The second point analyzed here is deeper. Generally covariant theories usually
have first class constraints quadratic in the momenta. Examples per excellence
of these kinds of theories are general relativity and string theory as well as
many toy models with finite degrees of freedom found in the literature. It
should be expected that their Hamiltonian actions were fully gauge-invariant
under the gauge symmetry. However, this is not so, rather, the gauge
transformed actions of these theories are equal to the original ones plus
boundary terms because of the Hamiltonian constraints quadratic in the
momenta. Therefore one usually handles gauge theories coming from nonfully
gauge-invariant actions which is an ugly situation from the point of view of
gauge theories. In this paper gauge symmetry is taken seriously and
variational principles with fully gauge-invariant actions are constructed by
adding suitable boundary terms to the non-gauge-invariant actions. By
introducing these boundary terms into the integral the quadratic constraints
are modified in a nice way: they become linear in the momenta (and homogeneous
in some cases) in the new canonical variables $({\cal Q}^i , {\cal P}_i)$
naturally induced by the boundary terms, which is a beautiful result just
coming from the full gauge invariance of their actions. Thus first class
constraints quadratic in the momenta are {\it not} the distinguishing mark of
generally covariant theories if fully gauge-invariant actions are taken into
account (cf. Ref. \cite{Teitelboim1982}). This result raises the question:
what then is a Hamiltonian constraint? If generally covariant systems endowed
with fully gauge-invariant actions and first class constraints quadratic in
the momenta can be written in terms of first class constraints linear in the
momenta like Yang-Mills theories, then how does one distinguish a genuine
Yang-Mills-like theory from one coming from a ``linearization" in the sense
explained of a generally covariant system? Does it make sense to distinguish
both types of theories just by the form that momenta enter in their
constraints? Even though it is not possible to give a definitive answer to
these questions from the present result, it is hoped that it can contribute to
make clear the meaning of Hamiltonian constraints in generally covariant
systems. In particular, the transformation of the constraints could be useful
to find in some cases a new set of solutions of physical states through the
Dirac condition $G_a({\cal Q}^i , {\cal P}_i) |\psi\rangle =0$. This
constitutes the second result of the present paper displayed in detail in Sec.
III. Finally, our conclusions are presented in Sec. IV.

Let us begin by recalling the canonical transformation induced when a boundary
term is present, because this is the heart of the methods used in this paper.
From now on, it is considered a generally covariant system determined by an
action of the form \cite{Dirac}
\begin{eqnarray}
S[q^i , p_i ,\lambda^a ] & = & \int_{\tau_1}^{\tau_2}d\tau (p_i
{\dot q}^i-H_E)-\left. B \right|^{\tau_2}_{\tau_1}\, , \quad
i=1,..., N \, ,\label{action0}
\end{eqnarray}
where $H_E = H_0 + \lambda^a G_a$ is the extended Hamiltonian, $G^a$ are first
class constraints, and $H_0$ is the canonical first class Hamiltonian,
\begin{eqnarray}
\{ G_a , G_b \} & = & C_{ab}\,^c G_c \, , \nonumber\\
\{ H_0, G_a \} & = & V_a \,^b G_b \, ,\quad a,b,c=1,...,M \, .
\end{eqnarray}
Therefore the system has $D=N-M$ continuous physical degrees of freedom ($2D$
in the reduced phase space). The boundary term $B(q,p,\tau)$ determines a
complete set of commuting variables $Q^i(q,p,\tau)$ fixed at the end points
$\tau_1$ and $\tau_2$,
\begin{eqnarray}
Q^i \left [ q(\tau_{\alpha}),p(\tau_{\alpha}),\tau_{\alpha} \right ] & = &
Q^i_{\alpha}
\, , \quad \alpha =1,2 \, ,\nonumber\\
\{ Q^i,Q^j\}& = & 0 \, , \quad \mbox{(at equal $\tau$'s)} \, .
\label{boundary0}
\end{eqnarray}
These commuting variables are equal in number as the $q$'s (``completeness")
and could be, for instance, the $q$'s themselves $(Q^i=q^i)$ or any other
combination of $q$'s and  $p$'s satisfying the commutation condition in the
Poisson brackets sense. The relationship between these variables and the
boundary term is given by
\begin{equation}
p_i\delta q^i - H_E \delta \tau-\delta B= P_i\delta Q^i - {\cal
H}_E \delta \tau \, ,
\end{equation}
where the $P$'s are the momenta conjugate to the $Q$'s, and ${\cal H}_E$ is
the new extended Hamiltonian. Taking independent variations of $q^i ,\ p_i$,
and $\tau$ yields
\begin{eqnarray}
p_i - {\partial B \over \partial q^i} - P_j {\partial Q^j \over
\partial q^i} & = & 0 \, , \nonumber\\
P_j{\partial Q^j \over \partial p_i} +
{\partial B \over \partial p_i} & = & 0 \, , \nonumber \\
{\cal H}_E -H_E -{\partial B \over \partial \tau}- {\partial Q^j
\over \partial \tau }P_j & = & 0 \, , \label{canonical}
\end{eqnarray}
which establish the relationship between the new phase space
variables $(Q^i , P_i )$ with the initial ones $(q^i , p_i)$.
There is no uniqueness in the solution of these equations, in a
similar way to the fact that a canonical transformation can be
obtained using different generating functions.

\section{Compatibility between boundary and gauge conditions using finite
gauge transformations}

{\it 1. Non-gauge-invariant actions}. It was stated in the Introduction that
this variational principle featuring finite gauge transformations is,
essentially, the finite version of the approach of Ref. \cite{Henneaux1992}
based on non-gauge-invariant actions. From a technical perspective the only
remark is that the interplay between first class constraints linear and
homogeneous in the momenta and quadratic (or higher) in the momenta could
imply a contribution to the boundary term of the former if finite gauge
transformations are taken into account [see the example of the $SL(2,R)$ model
in this section]. This is a key difference between the infinitesimal and
finite versions because in the infinitesimal approach constraints linear and
homogeneous in the momenta never contribute to boundary terms.

To be precise, this method is an analogue to the one in Ref.
\cite{Henneaux1992}. There, authors use an ``hybrid" variational principle
because the action and the boundary conditions are expressed in terms of the
original set of variables, however, the right gauge condition is expressed in
terms of the gauge related ones. Here, on the other hand, the standard
procedure is followed, namely, the action, the boundary conditions, and the
gauge condition are written in terms of the {\it same} set of canonical
variables: the gauge related ones. Before going to the description of the
method, boundary conditions (\ref{boundary0}) and algebraic gauge conditions
$\chi_a(q^i,p_i)=0$ are assumed not compatible to each other. Then, using the
finite gauge symmetry
\begin{eqnarray}
{q'}^i & = & {q'}^i (q^i,p_i, \alpha^a)\, , \quad {p'}_i = {p'}_i
(q^i,p_i,\alpha^a)\, , \label{gt}
\end{eqnarray}
generated by the first class constraints and the finite transformation law for
the Lagrange multipliers, the gauge transformed action (\ref{action0}) is
computed,
\begin{eqnarray}
S[{q'}^i, {p'}_i, {\lambda'}^a] & = & S[q^i,p_i,\lambda^a] +
\Delta S (q^i, p_i, \alpha^a)\mid^{\tau_2}_{\tau_1}\, .\label{FGT}
\end{eqnarray}
Therefore, the original action (\ref{action0}) can be expressed in
terms of the gauge related variables and the gauge parameters
\begin{eqnarray}
S[q^i,p_i, \lambda^a] & = & S[{q'}^i, {p'}_i,{\lambda'}^a] - \Delta S \left [
q^i({q'}^i,{p'}_i,\alpha^a), q_i({q'}^i,{p'}_i,\alpha^a),\alpha^a \right
]\mid^{\tau_2}_{\tau_1}\, , \label{JJJ}
\end{eqnarray}
obtained by plugging into the boundary term $\Delta S$ of Eq. (\ref{FGT}) the
inverse transformation of Eq. (\ref{gt}). At the same time, the original
boundary conditions (\ref{boundary0}) must be rewritten by inserting into the
left-hand side (LHS) of Eq. (\ref{boundary0}) the inverse transformation of
Eq. (\ref{gt}),
\begin{eqnarray}
Q^i \left [ q^i({q'}^i,{p'}_i,\alpha^a), p_i({q'}^i,{p'}_i,\alpha^a),\tau
\right ] (\tau_{\alpha}) &= & Q^i_{\alpha} \, , \quad \alpha=1,2. \label{ZZZ}
\end{eqnarray}
In this way, the variational principle defined in terms of the
gauge related variables, formed with the action (\ref{JJJ}) and
the boundary conditions (\ref{ZZZ}), has a certain freedom encoded
in the gauge parameters $\alpha^a$. The goal is to use this
freedom to make compatible both boundary and gauge conditions. In
the gauge related variables, the gauge condition is
\begin{eqnarray}
\chi_a ({q'}^i,{p'}_i)= 0 & \, . \label{GenericG}
\end{eqnarray}
Inserting Eq. (\ref{gt}) into Eq. (\ref{GenericG}) fixes the gauge parameters
\begin{eqnarray}
\alpha^a & = & \alpha^a (q^i,p_i)\, . \label{GenericFP}
\end{eqnarray}
The remaining task is to plug Eq. (\ref{GenericFP}) into the right-hand side
(RHS) of Eq. (\ref{JJJ}) and the LHS of Eq. (\ref{ZZZ}) to obtain a
well-defined variational principle with both boundary and gauge conditions
compatible to each other. Like in the infinitesimal case, $q^i(\tau_{\alpha})$
and $p_i(\tau_{\alpha})$ play the role of parameters. Of course, the
transversality condition has been assumed, which is not necessary if
noncanonical gauges are taken into account.

{\it 2. Fully gauge-invariant actions}. The detailed description of this
second type of variational principle, introduced here for the first time, both
in its finite and infinitesimal versions follows.

(i) Finite case. Knowing that the action (\ref{action0}) of any generally
covariant system transforms in the generic case like Eq. (\ref{FGT}), the goal
is to build a new action $S_{\mbox{\small inv}}$ such that it transforms like
\begin{eqnarray}\label{finvac}
S_{\mbox{\small inv}}[{q'}^i, {p'}_i, {\lambda'}^a] & \approx &
S_{\mbox{\small inv}}[q^i, p_i, \lambda^a]\, ,
\end{eqnarray}
under the finite gauge transformation. The weak expression (\ref{finvac})
means that the gauge invariance of the action is required only until a
boundary term proportional to the constraints. It is proposed for
$S_{\mbox{\small inv}}$ the form
\begin{eqnarray}
S_{\mbox{\small inv}}[q^i,p_i,\lambda^a] & \approx &
S[q^i,p_i,\lambda^a]-F(q^i,p_i) \mid^{\tau_2}_{\tau_1}=S_{\mbox{\small
inv}}[{\cal Q}^i,{\cal P}_i,\lambda^a] \, ,\label{GEN}
\end{eqnarray}
where $F$ is a function of the canonical variables and the last equality
is useful to remark that the action (\ref{GEN}) is subject
 to the boundary conditions
\begin{eqnarray}
{\cal Q}^i(q^i,p_i,\tau)(\tau_{\alpha}) & = & {\cal Q}^i_{\alpha}\, ,
\label{YYY}
\end{eqnarray}
determined by Eq. (\ref{canonical}) but with $B$ replaced by $B+F$. Of course,
it is always possible to add an arbitrary function of Dirac observables
$O_{\mbox{\small phys}}$, $\{O_{\mbox{\small phys}}, G_a \}=O_a\,^b
O_{\mbox{\small phys}}$ to the time boundary in the RHS of Eq. (\ref{GEN})
which would allow the introduction of {\it ad hoc} boundary conditions.

If the gauge condition is not compatible with Eq. (\ref{YYY}), then it can be
imposed in the gauge related variables (\ref{GenericG}), fixing the gauge
parameters to Eq. (\ref{GenericFP}). Again, the variational principle is
defined in terms of the gauge related variables, its action is simply Eq.
({\ref{GEN}) but written in terms of the gauge related canonical variables,
under the boundary conditions
\begin{eqnarray}
{\cal Q}^i({q'}^i, {p'}_i,\tau)(\tau_{\alpha}) & = & f^i \left [ {\cal
Q}_{\alpha}^i, q^i (\tau_{\alpha}) , p_i (\tau_{\alpha}),\tau_{\alpha} \right
] \, , \label{YYYII}
\end{eqnarray}
obtained by inserting into the LHS of Eq. (\ref{YYY}) both Eq.
(\ref{GenericFP}) and the inverse gauge transformation of Eq. (\ref{gt}).
Notice that it is not necessary to insert Eq. (\ref{GenericFP}) into Eq.
(\ref{GEN}) because such action is already gauge-invariant. This is a key
difference between the variational principle with non-gauge-invariant actions
and the current one. Thus the fixed gauge parameters (\ref{GenericFP}) are
inserted just into the boundary conditions. Moreover, Eqs. (\ref{YYY}) and
(\ref{YYYII}) have the same functional form in their LHS, the only difference
between them is the value they take in their RHS. Again, $q^i(\tau_{\alpha})$,
and $p_i(\tau_{\alpha})$ play the role of parameters. Finally, it is not
possible to determine the explicit form for $F$ in the generic case. However,
it is possible to write down the differential equations that $F$ must satisfy
in the infinitesimal approach.

(ii) Infinitesimal case. Knowing that Eq. (\ref{action0}) transforms
infinitesimally like
\begin{eqnarray}
S'[{q'}^i, {p'}_i, {\lambda'}^a] & = & S[q^i,p_i,\lambda^a]+ \left
(  p_i \frac{\partial G}{\partial p_i} - G  - \{ B, G \} \right
)\mid^{\tau_2}_{\tau_1}\, ,
\end{eqnarray}
and $F$ transforms as $F' =F + \{F,G\}$, then $F$ must weakly satisfy the set
of differential equations
\begin{eqnarray}
p_i \frac{\partial G}{\partial p_i} & \approx & \{ B+F
,G \} \, , \label{DIFFEQ}
\end{eqnarray}
with $G=\epsilon^a G_a$, $\epsilon^a$ being the infinitesimal gauge
parameters, then in Eq. (\ref{DIFFEQ}) there are $M$ differential equations
for $F$. Notice the interplay between $F$ and $B$, in particular $B=0$ when
the configuration variables $q^i$ are fixed at the time boundary ($\tau_1$ and
$\tau_2$) in variational principles with non-gauge-invariant actions. In this
case, $F$ is completely determined by the explicit form of the first class
constraints. If some of them were linear and homogeneous in the momenta then
they would not contribute to $F$, as it happens in Yang-Mills theories.
However, if the constraints are quadratic (or higher) in the momenta they will
contribute to $F$. Therefore it is only possible to build fully
gauge-invariant actions at least infinitesimally by solving Eq. (\ref{DIFFEQ})
and is not necessary to add new variables (enlarging the phase space) to build
fully gauge-invariant actions, as it is argued in Ref. \cite{Momen1997}. The
present formalism is completely systematic and could be easily extended to
field theory with the obvious modifications and compared with the known
results for self-dual gravity \cite{MonVer00} and bosonic string theory
\cite{MonReport} on this direction. Finally, the issues concerning the
transversality condition are those of Ref. \cite{Henneaux1992} for the same
reasons that apply in that case.

\subsection{Parametrized relativistic free particle}
\subsubsection{Non-gauge-invariant action}
The action for a parametrized relativistic free particle is
\begin{eqnarray}
S[x^{\mu} , p_{\mu}, \lambda ] & = & \int^{s_2}_{s_1} ds \left [
\frac{d x^{\mu}}{d s} p_{\mu} - \lambda \left ( p^{\mu} p_{\mu} +
m^2 c^2 \right ) \right ]\, , \label{FPaction}
\end{eqnarray}
where the inner product is taken with respect to the Minkowski metric
$\eta_{\mu\nu}$ with $\mbox{diag} \,\,\eta_{\mu \nu} = (-1,1,1,1)$,
$\mu=0,1,2,3$, $s$ is an arbitrary parameter (not necessarily the proper time
$\tau$) which parametrizes its world line. The standard boundary conditions
for Eq. (\ref{FPaction}) are
\begin{eqnarray}
x^{\mu} (s_{\alpha}) & = & x^{\mu}_{\alpha} \, , \quad \alpha =1,2 \, ,
\label{FPbound}
\end{eqnarray}
with $x^{\mu}_{\alpha}$ prescribed numbers. The constraint
\begin{eqnarray}
G := p^{\mu} p_{\mu} + m^2 c^2 = 0
\end{eqnarray}
is first class and generates a finite gauge transformation on the
phase space variables
\begin{eqnarray}
{x'}^{\mu} & = & x^{\mu} + \theta (s) p^{\mu}\, , \quad
p'_{\mu}=p_{\mu}\, , \label{FPtransI}
\end{eqnarray}
whereas the Lagrange multiplier transforms as
\begin{eqnarray}
\lambda' & = & \lambda + \frac12 \frac{d \theta}{d s} \label{FPtransII}
\end{eqnarray}
with $\theta(s)$ an arbitrary function of $s$. Now suppose the gauge
condition
\begin{eqnarray}
x^0 & = & 0  \label{FPgauge}
\end{eqnarray}
were picked. This gauge condition did not match the boundary conditions
(\ref{FPbound}) if $x^0_{\alpha} \neq 0$ and then the dynamical problem would
be in trouble. This conflict can be solved applying the general scheme already
explained. To do this, note that under the gauge symmetry given by Eqs.
(\ref{FPtransI}), (\ref{FPtransII}) the action (\ref{FPaction}) transforms
like
\begin{eqnarray}
S [{x'}^{\mu}, {p'}_{\mu}, \lambda' ] & = & S [x^{\mu}, p_{\mu} ,
\lambda ] + \left [ \left ( p^{\mu} p_{\mu} - m^2 c^2 \right )
\frac{\theta}{2} \right ]^{s_2}_{s_1} \, .
\end{eqnarray}
The boundary term $\Delta S =\left ( p^{\mu} p_{\mu} - m^2 c^2 \right )
({\theta}/{2})$ comes from the fact that $G$ is quadratic in the momenta.
Notice that $\Delta S$ is {\it not} proportional to $G$ and so $S$ is not
gauge-invariant even on the constraint surface.\footnote{In complex general
relativity expressed in terms of Ashtekar variables the boundary term is
proportional to the Hamiltonian constraint. See Ref. \cite{MonVer00} for the
details.} Thus the original action (\ref{FPaction}) can be expressed in terms
of the gauge related variables $({x'}^{\mu}, {p'}_{\mu})$, the Lagrange
multiplier $\lambda'$, and the gauge parameter $\theta$,
\begin{eqnarray}
S [x^{\mu}, p_{\mu}, \lambda ] & = & S [{x'}^{\mu}, {p'}_{\mu} ,
\lambda' ] - \left [ \left ( {p'}^{\mu} {p'}_{\mu} - m^2 c^2
\right ) \frac{\theta}{2} \right ]^{s_2}_{s_1} \, .
\label{FPactionII}
\end{eqnarray}
The RHS of the last expression tells $S [x^{\mu}, p_{\mu}, \lambda ]$ can be
considered a functional of the gauge related variables ${x'}^{\mu},
{p'}_{\mu}, \lambda'$ , and $\theta$. Also, the original boundary conditions
(\ref{FPbound}) must be written in terms of the gauge related variables
\begin{eqnarray}
{x'}^{\mu} (s_{\alpha}) - \theta (s_{\alpha}) p^{\mu} (s_{\alpha}) & = &
x^{\mu}_{\alpha} \, . \label{FPboundII}
\end{eqnarray}
It is time to set the right dynamical problem with boundary
conditions compatible with the gauge condition. In the gauge
related variables, the gauge condition is
\begin{eqnarray}
{x'}^0 & = & 0\, . \label{FPgaugeII}
\end{eqnarray}
By combining the gauge condition and Eq. (\ref{FPtransI}) the gauge parameter
is fixed,
\begin{eqnarray}
\theta & = & - \frac{x^0}{p^0}\, .
\end{eqnarray}
The goal has been reached. The remaining task is to plug into the RHS of Eq.
(\ref{FPactionII}) and into the LHS of Eq. (\ref{FPboundII}) last expression
for the gauge parameter. By doing this the right action,
\begin{eqnarray}
S_{\mbox{\small red}} [{x'}^{\mu}, {p'}_{\mu}, \lambda' ] (x^0_{\alpha},
p^0_{\alpha}) & : = & S [ x^{\mu}, p_{\mu} , \lambda ]
\mid_{\theta=-({x^0}/{p^0})} \, ,
\nonumber\\
& = & S [{x'}^{\mu}, {p'}_{\mu} , \lambda' ] + \left [ \left (
{p'}^{\mu} {p'}_{\mu} - m^2 c^2 \right ) \frac{1}{2}
\frac{x^0_2}{p^0_2} \right ] \nonumber\\
& & - \left [ \left ( {p'}^{\mu} {p'}_{\mu} - m^2 c^2 \right ) \frac{1}{2}
\frac{x^0_1}{p^0_1} \right ]\, , \label{FPactionIII}
\end{eqnarray}
is obtained, under the boundary conditions
\begin{eqnarray}
{x'}^{\mu} (s_{\alpha}) + \frac{x^0_{\alpha}}{p^0_{\alpha}} p^{\mu}_{\alpha} &
= & x^{\mu}_{\alpha} \, , \label{FPboundIII}
\end{eqnarray}
which are, by construction, compatible with the gauge condition
(\ref{FPgaugeII}). Here $x^{\mu}(s_{\alpha}) = x^{\mu}_{\alpha}$, and
$p^{\mu}(s_{\alpha}) = p^{\mu}_{ \alpha}$ which play the role of ``parameters"
in the final action (\ref{FPactionIII}).\footnote{From now on Kuchar's
notation is used, namely, $S[\cdots](\cdots)$ is a functional of the variables
inside the square brackets and a function of the variables inside the
parentheses.} In summary, the analysis began with a wrong variational
principle where the boundary conditions were not compatible with the gauge
condition, and a right variational principle with boundary conditions
compatible with the gauge condition was built. In the new variational
principle: (i) a new action including a boundary term was constructed [see Eq.
(\ref{FPactionIII})]; (ii) the boundary conditions were also modified [see Eq.
(\ref{FPboundIII})].

\subsubsection{Fully gauge-invariant action}
Now, the original action will be fully gauge-invariant under the gauge
symmetry generated by the constraint $G$ {\it in spite of the fact} the
constraint $G$ is quadratic in the momenta. The simplest boundary term needed
to build $S_{\mbox{\small inv}}$ is $F=x^{\mu} p_{\mu}$,
\begin{eqnarray}
S_{\mbox{\small inv}} [x^{\mu} , p_{\mu}, \lambda ] & = & \int^{s_2}_{s_1} ds
\left [ \frac{d x^{\mu}}{d s} p_{\mu} - \lambda \left ( p^{\mu} p_{\mu} + m^2
c^2 \right ) \right ] - x^{\mu} p_{\mu} \mid ^{s_2}_{s_1} \, .
\label{FPGIaction}
\end{eqnarray}
In fact, by using the finite gauge transformation (\ref{FPtransI}) and
(\ref{FPtransII}),
\begin{eqnarray}
S_{\mbox{\small inv}} [{x'}^{\mu}, {p'}_{\mu}, \lambda' ] & = &
S_{\mbox{\small inv}} [x^{\mu}, p_{\mu}, \lambda ] - \frac{\theta}{2} \left (
p^{\mu} p_{\mu} + m^2 c^2 \right ) \mid^{s_2}_{s_1}\, ,
\end{eqnarray}
and the difference between $S_{\mbox{\small inv}} [{x'}^{\mu}, {p'}_{\mu},
\lambda' ]$ and $ S_{\mbox{\small inv}} [x^{\mu}, p_{\mu}, \lambda ]$ is a
boundary term which is {\it proportional} to the first class constraint $G$.
Therefore $S_{inv}$ is gauge-invariant on the constraint surface $G=0$ only.
Of course an arbitrary function of the Dirac observables for the system
$F_1(p_{\mu}, x^{\mu} p^{\nu}-x^{\nu}p^{\mu})$ might have been (and can be)
added to the time boundary of Eq. (\ref{FPGIaction}) too without destroying
the gauge invariance of $S_{\mbox{\small inv}}$, just modifying the boundary
conditions.

To find the new boundary conditions associated with the action
(\ref{FPGIaction}) the canonical transformation induced by its boundary term
will be used. By using Eq. (\ref{canonical}) the new phase space variables,
\begin{eqnarray}
{\cal Q}^{\mu} & = & -\frac{1}{\beta} p^{\mu} \, , \quad
 {\cal P}_{\mu} = \beta x_{\mu} \, , \label{VFP}
\end{eqnarray}
are obtained, with $\beta$ a nonvanishing constant. Thus the boundary
conditions associated with Eq. (\ref{FPGIaction}) are
\begin{eqnarray}
{\cal Q}^\mu(s_\alpha)=-\frac{1}{\beta} p^{\mu} (s_{\alpha}) & = &
{\cal Q}^{\mu}_{\alpha} \, .
\label{XXX}
\end{eqnarray}
It is immediately seen that the wanted gauge condition $x^0 =0$ does not
conflict with these boundary conditions.

\subsection{Parametrized harmonic oscillator}
\subsubsection{Non-gauge-invariant action} In this case, the
original variational principle is defined by the action
\begin{eqnarray}
S[x, t, p, p_t , \lambda ] & = & \int^{\tau_2}_{\tau_1} d \tau
\left [ \frac{d x}{d \tau} p + \frac{d t}{d \tau} p_t - \lambda
\left ( p_t + \frac{p^2}{2m} + \frac12 m \omega^2 x^2 \right )
\right ] \, , \label{OA}
\end{eqnarray}
under the standard boundary conditions
\begin{eqnarray}
x (\tau_{\alpha}) & = & x_{\alpha} \, , \quad t (\tau_{\alpha}) = t_{\alpha}
\, , \quad \alpha=1,2 \, , \label{OBC}
\end{eqnarray}
with $x_{\alpha}$ and $t_{\alpha}$ prescribed numbers. The constraint
\begin{eqnarray}
G := p_t + \frac{p^2}{2m} + \frac12 m \omega^2 x^2 = 0\, ,
\end{eqnarray}
is first class and generates a {\it finite} gauge transformation on the phase
space variables \cite{MonGRG},
\begin{eqnarray}
x' & = & x \cos{\theta(\tau)}  +
\frac{p}{m \omega} \sin{\theta (\tau)} \, ,\nonumber\\
p' & = & -m \omega x \sin{\theta(\tau)} +
p \cos{\theta (\tau)} \, ,\nonumber\\
t'  & = & \frac{\theta (\tau)}{\omega} + t \, , \nonumber\\
{p'}_t & = & p_t \, , \label{OATRANS}
\end{eqnarray}
whereas the Lagrange multiplier transforms as
\begin{eqnarray}
\lambda' & = & \lambda + \frac{{\dot \theta} (\tau)}{\omega} \, ,
\label{OAlambda}
\end{eqnarray}
with ${\dot \theta} (\tau)= {d \theta (\tau)}/{d \tau}$.

Again, suppose the gauge condition
\begin{eqnarray} t= 0
\label{FG}
\end{eqnarray}
were picked. Obviously, it did not match the boundary conditions (\ref{OBC})
if $t_{\alpha} \neq 0$. It is time to apply the method. By using the {\it
finite} gauge transformation (\ref{OATRANS}), (\ref{OAlambda}) the action
(\ref{OA}) transforms like
\begin{eqnarray}
S[ x' , t' , p' , {p'}_t , \lambda' ] & = & S[x, t, p, p_t ,
\lambda ] + \left [ -\sin^2{\theta} x p + \frac{1}{\omega} \left (
\frac{p^2}{2m} - \frac{1}{2} m \omega^2 x^2 \right )
\frac{\sin{2\theta}}{2} \right ]^{\tau_2}_{\tau_1} \, .
\label{NonInv}
\end{eqnarray}
Again, the boundary term $\Delta S= -\sin^2{\theta} \,\, x p + (1/\omega)
\left ( \frac{p^2}{2m} - \frac{1}{2} m \omega^2 x^2 \right
)({\sin{2\theta})}/{2}$ comes from the quadratic in the momenta term of $G$.
Note that $\Delta S$ is {\it not} proportional to $G$. Therefore
\begin{eqnarray}
S[x, t, p, p_t , \lambda ] & = & S[ x' , t' , p' , {p'}_t ,
\lambda' ] - \left [ \sin^2{\theta} x' p' + \frac{1}{\omega} \left
( \frac{{p'}^2}{2m} - \frac{1}{2} m \omega^2 {x'}^2 \right )
\frac{\sin{2\theta}}{2} \right ]^{\tau_2}_{\tau_1} \, .\label{QQQ}
\end{eqnarray}
From the RHS of the last equation it is clear that $S[x, t, p, p_t , \lambda
]$ can be considered a functional of $x'$, $t'$, $p'$, ${p'}_t$, $\lambda'$,
and $\theta$. At the same time, the {\it original} boundary conditions
(\ref{OBC}) must be written in terms of the gauge related variables and the
gauge parameter $\theta$,
\begin{eqnarray}
x' (\tau_{\alpha}) \cos{\theta(\tau_{\alpha})} -
\frac{p'(\tau_{\alpha})}{m\omega} \sin{\theta(\tau_{\alpha})}
& = & x_{\alpha} \, ,\nonumber\\
t' (\tau_{\alpha}) - \frac{\theta(\tau_{\alpha})}{\omega} & = & t_{\alpha} \,
, \quad \alpha=1,2 \, . \label{OBCII}
\end{eqnarray}
It is time to define the new variational principle whose boundary
conditions will be compatible with the required gauge condition.
In the gauge related variables, the gauge condition is
\begin{eqnarray}
t' = 0 \, . \label{frozen}
\end{eqnarray}
By using the transformation law for the $t$ variable (\ref{OATRANS}) together
with the required gauge condition (\ref{frozen}) the explicit expression for
the gauge parameter is obtained, $\theta= -\omega t$. The goal has been
reached. A right variational principle with canonical pairs $(x',p')$ and $(t'
, {p'}_t )$ can be built, its action is \eject \vfill
\begin{eqnarray}
S_{\mbox{\small red}} [x' ,t' , p' , {p'}_t, \lambda'] (t_1 , t_2) & := &
S [x, t , p , p_t , \lambda ] \mid_{\theta=-\omega t} \, , \nonumber\\
& = & S[ x' , t' , p' , {p'}_t , \lambda' ] \nonumber\\
& & - \left [ \sin^2{\omega t_2} \,\, x' (\tau_2) p' (\tau_2) -
\frac{1}{\omega} \left ( \frac{{p'}^2 (\tau_2)}{2m} - \frac{1}{2} m \omega^2
{x'}^2 (\tau_2) \right ) \frac{\sin{2\omega t_2}}{2}
\right ]  \nonumber\\
& & + \left [ \sin^2{\omega t_1} \,\, x' (\tau_1) p' (\tau_1) -
\frac{1}{\omega} \left ( \frac{{p'}^2 (\tau_1) }{2m} - \frac{1}{2} m \omega^2
{x'}^2 (\tau_1) \right ) \frac{\sin{2\omega t_1}}{2}
\right ] \, , \nonumber\\
& & \label{modified}
\end{eqnarray}
under the boundary conditions
\begin{eqnarray}
x' (\tau_{\alpha}) \cos{\omega t_{\alpha}} + \frac{p'(\tau_{\alpha})}{m\omega}
\sin{\omega t_{\alpha}}
& = & x_{\alpha} \, ,\nonumber\\
t' (\tau_{\alpha}) & = & 0 \, , \quad \alpha=1,2 \, ,
\label{BCmodified}
\end{eqnarray}
which are, by construction, compatible with the gauge condition
\begin{eqnarray}
t' & = & 0 \, .
\end{eqnarray}
Once the dynamical problem has been well defined, there are still two
remaining things to do. The first one is to compute the gauge-fixed
variational principle by plugging into the action $S_{\mbox{\small red}} [x'
,t' , p' , {p'}_t, \lambda'] (t_1 , t_2)$ and into the boundary conditions
(\ref{BCmodified}) the gauge condition $t'=0$ as well as the constraint
$G'=0$. The second one is to solve the dynamics by using the equations of
motion with the gauge condition $t'=0$. Let us focus in the first option. By
plugging the gauge condition $t'=0$ and the constraint $G'=0$ into the action
$S_{\mbox{\small red}} [x' ,t' , p' , {p'}_t, \lambda'] (t_1 , t_2)$, the
gauge-fixed action
\begin{eqnarray}
S_{\mbox{\small fixed}} [x', p'](t_1 , t_2 ) & = & S_{\mbox{\small red}} [x ,t
, p' , {p'}_t, \lambda'] (t_1 , t_2) \mid_{G'=0, t'=0}
\, ,\nonumber\\
& = &
S [x, t , p , p_t , \lambda ]_{\theta=-\omega t, G'=0, t'=0} \, , \nonumber\\
& = & \int^{\tau_2}_{\tau_1} d \tau \frac{dx'}{d \tau} p'
\nonumber\\
& & - \left [ \sin^2{\omega t_2} \,\, x' (\tau_2) p'(\tau_2) -
\frac{1}{\omega} \left ( \frac{{p'}^2 (\tau_2) }{2m} - \frac{1}{2} m \omega^2
{x'}^2 (\tau_2) \right ) \frac{\sin{2\omega t_2}}{2}
\right ]  \nonumber\\
& & + \left [ \sin^2{\omega t_1} \,\, x' (\tau_1) p' (\tau_1) -
\frac{1}{\omega} \left ( \frac{{p'}^2 (\tau_1)}{2m} - \frac{1}{2} m \omega^2
{x'}^2 (\tau_1) \right ) \frac{\sin{2\omega t_1}}{2} \right ] \, .
\label{OAHHI}
\end{eqnarray}
is obtained. This form of the action is very interesting. It contains a
kinematical term like any action, but it has not a Hamiltonian, rather, all
its dynamical information has been mapped to its time boundary. Therefore it
is natural to interpret this result as the ``canonical version" of the
holographic hypothesis in the sense that with this particular choice of the
gauge condition, its dynamics is now at the time boundary \cite{Susskind1995}.
In fact, there was a ``reduction" of degrees of freedom, before fixing the
gauge the initial dynamical problem was defined on the closed set $[\tau_1 ,
\tau_2]$ whereas the final dynamical problem is now defined on just {\it two}
points, $\tau_2$ and $\tau_1$. Of course, the gauge-fixed variational
principle has associated the remaining boundary conditions,
\begin{eqnarray}
x' (\tau_{\alpha}) \cos{\omega t_{\alpha}} + \frac{p'(\tau_{\alpha})}{m\omega}
\sin{\omega t_{\alpha}} & = & x_{\alpha} \, , \quad \alpha = 1,2 \, .
\end{eqnarray}
Note that the number of boundary conditions has decreased.

Alternatively, the second option is to solve the equations of motion which are
the original ones but with $x,t,p,p_t$, and $\lambda$ replaced by $x',t', p',
{p'}_t$, and $\lambda'$. Their solution, using the gauge $t'=0$, is
\begin{eqnarray}
x' & = & x_0 \, , \nonumber\\
p' & = & p_0 \, ,\nonumber\\
t' & = & 0 \, , \nonumber\\
{p'}_t & = & - \frac{p^2_0}{2m} - \frac{1}{2}m \omega^2 x^2_0 \, ,
\label{Solution}
\end{eqnarray}
with $x_0$, $p_0$ constants ($\tau$ independent), and therefore they are Dirac
observables. Inserting this solution into the action (\ref{modified}), the
term $S[ x' , t' , p' , {p'}_t , \lambda' ]$ vanishes, and the only
contribution is given by the boundary term
\begin{eqnarray}
S (x_0 , p_0; t_1, t_2 ) & = & - \left [ \sin^2{\omega t_2}\,\, x_0 p_0 -
\frac{1}{\omega} \left ( \frac{{p_0}^2}{2m} - \frac{1}{2} m \omega^2 {x_0}^2
\right ) \frac{\sin{2\omega t_2}}{2}
\right ] \nonumber\\
& & + \left [ \sin^2{\omega t_1}\,\, x_0 p_0 - \frac{1}{\omega} \left (
\frac{{p_0}^2}{2m} - \frac{1}{2} m \omega^2 {x_0}^2 \right )
\frac{\sin{2\omega t_1}}{2} \right ] \, , \label{OAHHII}
\end{eqnarray}
where $x' (\tau_{\alpha})=x_0$ and $p' (\tau_{\alpha})=p_0$ were used. It is
clear that $S(x_0 , p_0; t_1, t_2)$ represents a two parameter family of
physical (Dirac) observables on the reduced phase space labeled by $x_0$ and
$p_0$; $t_1$ and $t_2$ being the parameters. Of course, Eqs. (\ref{OAHHI}) and
(\ref{OAHHII}) are the same thing but they look different because it has not
been inserted in Eq. (\ref{OAHHI}) the fact-coming from the equations of
motion-that $x' = x_0$ and $p' = p_0$. Using this information, Eq.
(\ref{OAHHI}) acquires the form (\ref{OAHHII}). In addition, the boundary
conditions, of course, reduce to
\begin{eqnarray}
x_0 \cos{\omega t_{\alpha}} + \frac{p_0 }{m\omega} \sin{\omega t_{\alpha}} & =
& x_{\alpha} \, , \quad \alpha=1,2 \, ,
\end{eqnarray}
establishing a relationship between the initial and final data [$x_1$, $x_2$,
$t_1$, and $t_2$] and the physical states ($x_0$, and $p_0$) in the reduced
phase space, displaying the fact that dynamics of the parametrized harmonic
oscillator between $\tau_1$ and $\tau_2$ is pure gauge, namely, it corresponds
to a point $(x_0,p_0)$ in the reduced phase space for each set of initial and
final data $(x_1, x_2, t_1, t_2)$.

To find the explicit expressions for these observables in terms of the
original phase space variables $(x,p)$, and $(t, p_t)$, it is necessary to use
the relationship between the original phase space variables $(x,t,p,p_t)$ and
the gauge related ones $(x',t',p',{p'}_t)$ together with the expression for
the gauge parameter $\theta=-\omega t$. From them,
\begin{eqnarray}
x' & = & x \cos{\omega t} -
\frac{p}{m \omega} \sin{\omega t} \, ,\nonumber\\
p' & = &  m \omega x \sin{\omega t} + p \cos{\omega t} \, .
\end{eqnarray}
But, because of Eq. (\ref{Solution}), these two expressions for $x'$ and $p'$
must be the same thing. Therefore $x_0$ and $p_0$ are given by \cite{MonGRG}
\begin{eqnarray}\label{x0p0}
x_0 & = & x \cos{\omega t} - \frac{p}{m \omega} \sin{\omega t} \,
,\nonumber\\
p_0 & = &  m \omega x \sin{\omega t} + p \cos{\omega t} \, ,
\end{eqnarray}
satisfying $\{ x_0 , p_0 \} =1$. Computing Eq. (\ref{x0p0}) at $t=0$, it can
be shown that these variables correspond to the initial conditions.
Furthermore, considering the inverse transformation of Eq. (\ref{x0p0}) it
follows that dynamics of the system is expressed in terms of the initial
conditions.

\subsubsection{Fully gauge-invariant action}
Now, the original action will be fully gauge-invariant under the gauge
symmetry generated by the first class constraint $G$ {\it in spite of the
fact} that the constraint $G$ is quadratic in the momentum $p$, ${p^2}/{2m}$.
The simplest boundary term needed to build $S_{\mbox{\small inv}}$ is
$F=-\frac12 xp$,
\begin{eqnarray}
S_{\mbox{\small inv}} [x, t, p, p_t , \lambda ] & = & \int^{\tau_2}_{\tau_1} d
\tau \left [ \frac{d x}{d \tau} p + \frac{d t}{d \tau} p_t - \lambda \left
(p_t + \frac{p^2}{2m} + \frac12 m \omega^2 x^2 \right ) \right ] - \frac{1}{2}
x p \mid^{\tau_2}_{\tau_1} \, . \label{OAInv}
\end{eqnarray}
Using Eqs. (\ref{OATRANS}) and (\ref{OAlambda}), it is clear that
\begin{eqnarray}
S_{\mbox{\small inv}} [x', t', p', {p'}_t , \lambda' ] & = & S_{\mbox{\small
inv}} [x, t, p, p_t , \lambda ] \, .
\end{eqnarray}
Thus the action $S_{\mbox{\small inv}}$ is indeed fully gauge-invariant under
the finite gauge symmetry involved. Again, an arbitrary function
$F_1({p^2}/{2m}+ \frac12 m\omega^2 x^2)$ might have been (and can be) added to
the time boundary of Eq. (\ref{OAInv}). However, Eq. (\ref{OAInv}) is the
simplest form. The boundary term in Eq. (\ref{OAInv}) induces the canonical
transformation (\ref{canonical}) from the original set of variables
$(x,t,p,p_t)$ to the new set $(\cal {X,T,P,P_T})$,
\begin{eqnarray}
{\cal X} & = & \frac12 \ln{\left ( \frac{x}{p}\right )}\, ,\quad {\cal P}=xp\, ,
\quad {\cal T}=t\,
,\quad {\cal P_T} =p_t\, . \label{WWW}
\end{eqnarray}
Therefore the boundary conditions associated with the action $S_{\mbox{\small
inv}}$ are
\begin{eqnarray}
\frac12 \ln{\left ( \frac{x}{p}\right )} (\tau_{\alpha}) & = & {\cal
X}_{\alpha} \, , \quad t (\tau_{\alpha} ) = {\cal T}_{\alpha} \, , \quad
\alpha=1,2 \, , \label{BC}
\end{eqnarray}
where a dimensional constant equals to $1$ is understood. Again, suppose the
gauge condition
\begin{eqnarray}
t = 0 \label{GC}
\end{eqnarray}
were picked, then it is pretty obvious that it would conflict the boundary
conditions (\ref{BC}) provided ${\cal T}_{\alpha} \neq 0$. Nevertheless, the
gauge condition can be reached in the gauge-related variables, namely,
\begin{eqnarray}
t' & = & 0\, . \label{NGC}
\end{eqnarray}
By applying the method, the right action is
\begin{eqnarray}
S_{\mbox{\small inv}} [x' , t' , p' , {p'}_t , \lambda'] & = &
\int^{\tau_2}_{\tau_1} d \tau \left [ \frac{d x'}{d \tau} p' + \frac{d t'}{d
\tau} {p'}_t - \lambda' \left ( {p'}_t + \frac{{p'}^2}{2m} + \frac12 m
\omega^2 {x'}^2 \right ) \right ]
\nonumber\\
& & - \frac{1}{2} x' p' \mid ^{\tau_2}_{\tau_1}\, ,
\end{eqnarray}
under the boundary conditions
\begin{eqnarray}
\frac12 \ln{\left ( \frac{x'}{p'}\right )} (\tau_{\alpha}) & = &
\frac{1}{2} \ln{\left ( \frac{e^{2 {\cal X}_{\alpha}} \cos{\omega
{\cal T}_{\alpha}} - \frac{1}{m\omega} \sin{\omega {\cal T}_{\alpha}}} {e^{2
{\cal X}_{\alpha}} \sin{\omega {\cal T}_{\alpha}} + \cos{\omega {\cal T}_{\alpha}}}
\right )}
\, ,\nonumber\\
t' (\tau_{\alpha}) & = & 0 \, , \quad \alpha =1,2 \, ,
\label{BCInvF}
\end{eqnarray}
which are, by construction, compatible with the gauge condition (\ref{NGC}).
Therefore this variational principle based on a fully gauge-invariant action
is more ``economic" than the one based on a non-gauge-invariant action because
in the former it is {\it not} necessary to handle additional boundary terms
for the original action is already fully gauge-invariant. Also, the boundary
conditions in terms of the original variables and in terms of the gauge
related ones look more ``symmetric" [see Eqs. (\ref{BC}) and (\ref{BCInvF})],
the difference between them being the value they take in their RHS.

Again, once the gauge conditions have been made compatible with the boundary
conditions, there are still two remaining things to do. The first one is to
compute the gauge-fixed variational principle by plugging into the action and
into the boundary conditions the gauge condition $t'=0$ together with the
constraint $G'=0$. By doing this
\begin{eqnarray}
S_{\mbox{\small fixed}} [x' , p'] & =&
S_{\mbox{\small inv}} [x' , t' , p' , {p'}_t , \lambda'] \mid_{G'=0 , t'=0} \, ,\nonumber\\
& = & \int^{\tau_2}_{\tau_1} d \tau \frac{d x'}{d \tau} p' -
\frac{1}{2} x' p' \mid ^{\tau_2}_{\tau_1}\, ,
\end{eqnarray}
is the gauge-fixed action and its boundary conditions are
\begin{eqnarray}
\frac12 \ln{\left ( \frac{x'}{p'}\right )} (\tau_{\alpha}) & = &
\frac{1}{2} \ln{\left ( \frac{e^{2 {\cal X}_{\alpha}} \cos{\omega
{\cal T}_{\alpha}} - \frac{1}{m\omega} \sin{\omega {\cal T}_{\alpha}}} {e^{2
{\cal X}_{\alpha}} \sin{\omega {\cal T}_{\alpha}} + \cos{\omega {\cal T}_{\alpha}}}
\right )} \, , \quad \alpha =1,2 .
\end{eqnarray}
This form of the action contains a kinetic term, as any action, and it has not
a Hamiltonian. Where is the dynamics contained? Obviously it is fully
contained in the boundary terms, in a similar way to Eq. (\ref{x0p0}). Again,
this result might be interpreted as an implementation of the holographic
hypothesis in the sense dynamics has been mapped to its time boundary
\cite{Susskind1995}.

Finally, it can be easily checked that the dynamics coming from the last
variational principle is the same as the one coming from the equations of
motion when the gauge condition $t'=0$ is imposed, as was done in the
noninvariant case discussed in the previous subsection. It makes no sense to
repeat this computation.

\subsection{$SL(2,R)$ model}

\subsubsection{Non-gauge-invariant $SL(2,R)$ model}
Up to here, generally covariant systems with a single Hamiltonian constraint
have been studied. Next, a model with two noncommuting Hamiltonian constraints
and one constraint linear and homogeneous in the momenta will be analyzed. The
nontrivial interplay among linear and quadratic constraints will produce a
contribution to the boundary term of the linear ones in direct opposition to
what the infinitesimal approach says.

This model has a $SL(2,R)$ gauge symmetry, one continuous physical degree of
freedom, and mimics the constraint structure
\begin{eqnarray}
\{ H , H \} \sim D \, , \quad \{ H , D \} \sim H \, , \quad \{ D ,
D\} \sim D \, , \quad \label{Algebra}
\end{eqnarray}
of general relativity. It can be considered as a (two points) discrete version
of the Arnowitt-Deser-Misner (ADM) formulation of gravity. Readers are urged
to read Ref. \cite{Mon99} for the details, in particular for a clear
description of the relational evolution of the degrees of freedom of the
system (on this see also Refs. \cite{Rovelli,MonGRG,Gambini01}). Its
Hamiltonian action is
\begin{eqnarray}
S [ {\vec u} , {\vec v} , {\vec p} , {\vec \pi}, N , M, \lambda ]
& = & \int_{\tau_1}^{\tau_2} d\tau \left [ \dot {\vec u} \cdot
{\vec p} + \dot {\vec v} \cdot {\vec \pi} - \left ( N H_1 + M H_2
+\lambda D \right ) \right ]\, , \label{action}
\end{eqnarray}
under the boundary conditions
\begin{eqnarray}
{\vec u} (\tau_{\alpha}) & = & {\vec U}_{\alpha} \, , \quad
 {\vec v} (\tau_{\alpha}) = {\vec V}_{\alpha} \, , \quad \alpha = 1,2
\, . \label{boundaryI}
\end{eqnarray}
The canonical pairs are $(\vec{u} ,\vec{p})$, and $(\vec{v} ,\vec{\pi})$; each
vector being a two-dimensional real one, the scalar product is taken in $E^2$.
$N$, $M$, and $\lambda$ are Lagrange multipliers. The constraints $H_1$,
$H_2$, and $D$,
\begin{eqnarray}
H_1 & := & \frac12 ({\vec p}^2 - {\vec v}^2)=0\, ,\nonumber\\
H_2 & := & \frac12 ( {\vec \pi}^2 - {\vec u}^2)=0\, ,\nonumber\\
D\ & := & {\vec u}\cdot {\vec p} - {\vec v} \cdot {\vec \pi}=0\, ,
\label{Constraints}
\end{eqnarray}
are first class, with ${\vec p}^2 = \vec p \cdot \vec p = (p_1)^2 + (p_2)^2$,
and so on. The constraint algebra is isomorphic to the $sl(2,R)$ Lie algebra
and the {\it finite} gauge transformation the constraints generate is
\cite{Mon99}
\begin{eqnarray}
{\vec u}'  =  \alpha(\tau) {\vec u} + \beta(\tau) {\vec p}\, ,
&\hspace{2em}& {\vec \pi}'  =  \alpha(\tau) {\vec \pi} +
\beta(\tau) {\vec v}\, ,
\nonumber\\
{\vec p}'  = \gamma(\tau) {\vec u} + \delta(\tau) {\vec p}\, ,
&\hspace{2em}& {\vec v}'  =  \gamma(\tau) {\vec \pi} +
\delta(\tau) {\vec v}\, , \label{Variables}
\end{eqnarray}
where the matrix
\begin{equation}
G(\tau) = \pmatrix{\alpha(\tau)  & \beta(\tau) \cr \gamma(\tau) & \delta(\tau)
}  \label{Gt}
\end{equation}
belongs to the $SL(2, R)$ group, i.e., it satisfies
$\alpha(\tau)\delta(\tau)-\beta(\tau)\gamma(\tau)=1\,$. So, the
system is invariant under a $SL(2, R)$ gauge symmetry local in
$\tau$. The finite transformation law for the Lagrange multipliers
is
\begin{eqnarray}
\pmatrix{ \lambda' & N' \cr M' & -\lambda'} & = & \pmatrix{ \alpha
& \beta \cr \gamma & \delta} \pmatrix{ \lambda & N \cr M &
-\lambda} \pmatrix{ \delta & -\beta \cr -\gamma & \alpha}
\nonumber\\
& & - \pmatrix{ \alpha & \beta \cr \gamma & \delta} \pmatrix{
\dot{\delta} & -\dot{\beta} \cr -\dot{\gamma} & \dot{\alpha}} \, ,
\label{Multipliers}
\end{eqnarray}
so, the matrix (\ref{Multipliers}) transforms as a Yang-Mills
connection valued in the Lie algebra of $SL(2,R)$ \cite{Mon99}.

Now, suppose the gauge condition
\begin{eqnarray}
u^1 & = & A \, ,\quad u^2 = B \, , \quad p_1 = C \label{ONIGC}
\end{eqnarray}
were picked. Obviously it did not match with the boundary
conditions (\ref{boundaryI}) in general. Under the {\it finite}
gauge transformation (\ref{Variables}) and (\ref{Multipliers}) the
change of the action (\ref{action}) is
\begin{eqnarray}
S [ {\vec u}' , {\vec v}' , {\vec p}' , {\vec \pi}', N' , M' , \lambda' ] & =
& S [ {\vec u} , {\vec v} , {\vec p} , {\vec \pi}, N , M, \lambda ]
\nonumber\\
& & + \left [ (\beta\gamma) ({\vec u}\cdot {\vec p} + {\vec v} \cdot {\vec
\pi} ) + \frac12 (\alpha\gamma) ({\vec u}^2 +{\vec \pi}^2) + \frac12
(\beta\delta) ({\vec v}^2 + {\vec p}^2) \right ]^{\tau_2}_{\tau_1}\, .
\label{transS}
\end{eqnarray}
The boundary term $\Delta S= (\beta\gamma) ({\vec u}\cdot {\vec p} + {\vec v}
\cdot {\vec \pi} ) + \frac12 (\alpha\gamma) ({\vec u}^2 +{\vec \pi}^2) +
\frac12 (\beta\delta) ({\vec v}^2 + {\vec p}^2)$ comes from the two
noncommuting Hamiltonian constraints. Also, $\Delta S$ is {\it not} a linear
combination of the first class constraints. More important, the term
$(\beta\gamma) ({\vec u}\cdot {\vec p} + {\vec v} \cdot {\vec \pi} )$ is {\it
not} present at the infinitesimal level. This term is associated with the
``diffeomorphism" constraint $D$, which is linear and homogeneous in the
momenta and so it does not contribute at the {\it infinitesimal} level.
Nevertheless, when {\it finite} gauge transformations are taken into account,
the contribution associated with this constraint appears again. Notice that if
the two Hamiltonian constraints were turned off, namely, $\beta =0$ and
$\gamma =0$, then the action would be gauge-invariant as expected because the
only remaining constraint would be $D$, which is linear and homogeneous in the
momenta. Here, it will be taken into account the full $SL(2,R)$ gauge symmetry
and not just only a subgroup of it. By using again the finite gauge
transformation (\ref{Variables}) and (\ref{Multipliers}) the original action
can be written as
\begin{eqnarray}
S [ {\vec u} , {\vec v} , {\vec p} , {\vec \pi}, N , M, \lambda ] & = & S [
{\vec u}' , {\vec v}' , {\vec p}' , {\vec \pi}', N' , M' , \lambda' ]
\nonumber\\
& & - \left [ -(\beta\gamma) ({\vec u}' \cdot {\vec p}' + {\vec v}' \cdot
{\vec \pi}' ) + \frac12 (\gamma\delta) ({\vec{u'}}^2 +{\vec{\pi'}}^2) +
\frac12 (\alpha\beta) ({\vec{v'}}^2 +
{\vec{p'}}^2) \right ]^{\tau_2}_{\tau_1}\, . \nonumber\\
& & \label{NGIaction}
\end{eqnarray}
From the RHS of last expression it is clear that $S [ {\vec u} , {\vec v} ,
{\vec p} , {\vec \pi}, N , M, \lambda ] $ can be interpreted as a functional
of ${\vec u}'$, ${\vec v}'$, ${\vec p}'$, ${\vec \pi}'$, $N'$, $M'$,
$\lambda'$, $\alpha$, $\beta$, $\gamma$, and $\delta$. In the same way, it is
possible to rewrite the {\it original} boundary conditions (\ref{boundaryI})
in terms of the gauge related variables as well as of the gauge parameters
$\alpha$, $\beta$, $\gamma$, and $\delta$,
\begin{eqnarray}
\delta (\tau_{\alpha}) {u'}^i (\tau_{\alpha}) - \beta
(\tau_{\alpha})
{p'}_i (\tau_{\alpha}) & = & U^i_{\alpha} \, ,\nonumber\\
\alpha (\tau_{\alpha}) {v'}^i (\tau_{\alpha}) - \gamma
(\tau_{\alpha}) {\pi'}_i (\tau_{\alpha}) & = & V^i_{\alpha} \, ,
\quad \alpha = 1,2; \quad i=1,2. \label{boundaryIN}
\end{eqnarray}
It is time to define the new variational principle whose boundary conditions
will be compatible with the required gauge condition. This new variational
principle is defined in terms of the gauge related variables $({\vec u}',
{\vec v}', {\vec p}', {\vec \pi}')$, given by the RHS side of Eq.
(\ref{NGIaction}), and its boundary conditions will be those given in Eq.
(\ref{boundaryIN}). In the gauge related variables, the gauge condition is
\begin{eqnarray}
{u'}^1 & = & A \, , \quad {u'}^2 = B \, , \quad {p'}_1 = C \, .
\label{NIGC}
\end{eqnarray}
The explicit expressions for the gauge parameters are computed using Eqs.
(\ref{Variables}), (\ref{NIGC}) together with $\alpha \delta -\beta \gamma=1$,
\begin{eqnarray}
\alpha & = & \frac{A p_2 - B p_1}{O_{12}}\, , \nonumber\\
\beta & = &  \frac{B u^1 - A u^2}{O_{12}}\, , \nonumber\\
\gamma & = & \frac{AC p_2 - (BC + O_{12})p_1}{A O_{12}} \, , \nonumber\\
\delta & = & \frac{-AC u^2 + (BC + O_{12}) u^1}{A O_{12}}\, ,
\label{EGP}
\end{eqnarray}
where $O_{12}=u^1 p_2 - u^2 p_1$ is a physical observable
\cite{Mon99}. The goal has been reached, i.e., it has been
possible to build a variational principle where the canonical
pairs are $({\vec u}', {\vec p}')$ and $({\vec v}' , {\vec
\pi}')$, its action is given by
\begin{eqnarray}
& & S_{\mbox{\small red}} [{\vec u}' , {\vec v}' , {\vec p}', {\vec \pi}', N'
, M' , \lambda'] (u^i (\tau_{\alpha}), p_i (\tau_{\alpha})) := S [{\vec u},
{\vec v} , {\vec p} , {\vec \pi} , N, M, \lambda ]
\mid_{\alpha, \beta, \gamma, \delta} \, , \nonumber\\
& = & S[ {\vec u}' , {\vec v}' , {\vec p}' , {\vec \pi}', N' , M'
, \lambda' ]  \nonumber\\
& & + \left [ \left ( \frac{B u^1 - A u^2}{O_{12}} \right ) \left
( \frac{AC p_2 - (BC + O_{12})p_1}{A O_{12}} \right ) ({\vec u}'
\cdot {\vec p}' + {\vec v}' \cdot {\vec \pi}' )
\right ]^{\tau_2}_{\tau_1} \nonumber\\
& & - \left [ \frac12 \left ( \frac{AC p_2 - (BC + O_{12})p_1}{A
O_{12}} \right ) \left ( \frac{-AC u^2 + (BC + O_{12}) u^1}{A
O_{12}} \right ) ({\vec{u'}}^2 +{\vec{\pi'}}^2)
\right ]^{\tau_2}_{\tau_1} \nonumber\\
& & -\left [ \frac12 \left ( \frac{A p_2 - B p_1}{O_{12}} \right ) \left (
\frac{B u^1 - A u^2}{O_{12}} \right ) ({\vec{v'}}^2 + {\vec{p'}}^2) \right ]
^{\tau_2}_{\tau_1} \label{NIAmod}
\end{eqnarray}
and the boundary conditions are
\begin{eqnarray}
\left ( \frac{-AC u^2 + (BC + O_{12}) u^1}{A O_{12}} \right )
(\tau_{\alpha}) \,\, u'^{i} (\tau_{\alpha}) - \left ( \frac{B u^1
- A u^2}{O_{12}} \right ) (\tau_{\alpha}) \,\, p'_{i}
(\tau_{\alpha}) & = & U^i_{\alpha} \, , \nonumber\\
- \left ( \frac{AC p_2 - (BC + O_{12})p_1}{A O_{12}} \right )
(\tau_{\alpha}) \,\, \pi_{i}' (\tau_{\alpha}) + \left ( \frac{A
p_2 - B p_1}{O_{12}} \right ) (\tau_{\alpha}) \,\, v'^{i}
(\tau_{\alpha}) & = & V^i_{\alpha} \, , \quad \alpha=1,2; \quad
i=1,2, \nonumber\\
& & \label{NIBCmod}
\end{eqnarray}
which are, by construction, compatible with the gauge condition
(\ref{NIGC}).

Once the dynamical problem has been well defined, there are still two
remaining things to do. The first one is to compute the gauge-fixed
variational principle by plugging into both the action $S_{\mbox{\small red}}
[{\vec u}' , {\vec v}' , {\vec p}', {\vec \pi}', N' , M' , \lambda'] \left (
u^i (\tau_{\alpha}), p_i (\tau_{\alpha}) \right )$ and into the boundary
conditions (\ref{NIBCmod}) the gauge condition (\ref{NIGC}) together with the
first class constraints equal to zero. The second one is to solve the dynamics
by using the equations of motion and the gauge condition (\ref{NIGC}). Let us
focus on the first option. By plugging the gauge condition (\ref{NIGC}) and
the first class constraints equal to zero into the action $S_{\mbox{\small
red}} [{\vec u}' , {\vec v}' , {\vec p}', {\vec \pi}', N' , M' , \lambda']
\left ( u^i (\tau_{\alpha}), p_i (\tau_{\alpha}) \right )$, the gauge-fixed
action
\begin{eqnarray}
& & S_{\mbox{\small fixed}}[{\vec v}' , {\vec \pi}'] \left ( u^i
(\tau_{\alpha}) \right ) =
\nonumber\\
& & = S_{\mbox{\small red}} [{\vec u}' , {\vec v}' , {\vec p}', {\vec \pi}',
N' , M' , \lambda'] \left ( u^i (\tau_{\alpha}), p_i (\tau_{\alpha}) \right )
\mid_{\alpha,\beta,\gamma,\delta; {H'}_1=0, {H'}_2=0, D'=0,
{u'}^1=A, {u'}^2 =B, {p'}_1 =C} \, , \nonumber\\
& = & \int^{\tau_2}_{\tau_1} d \tau \left [ \frac{d {v'}^1}{d\tau}
{\pi'}_1 +
\frac{d {v'}^2}{d\tau} {\pi'}_2 \right ] \nonumber\\
& & + \left [ \left ( \frac{B u^1 - A u^2}{O_{12}} \right ) \left
( \frac{AC p_2 - (BC + O_{12})p_1}{A O_{12}} \right ) ( 2 {\vec
v}' \cdot {\vec \pi}' )
\right ]^{\tau_2}_{\tau_1} \nonumber\\
& & - \left [ \left ( \frac{AC p_2 - (BC + O_{12})p_1}{A O_{12}}
\right ) \left ( \frac{-AC u^2 + (BC + O_{12}) u^1}{A O_{12}}
\right ) {\vec{\pi'}}^2
\right ]^{\tau_2}_{\tau_1} \nonumber\\
& & -\left [ \left ( \frac{A p_2 - B p_1}{O_{12}} \right ) \left (
\frac{B u^1 - A u^2}{O_{12}} \right ) {\vec{v'}}^2  \right ]
^{\tau_2}_{\tau_1}\, ,
\end{eqnarray}
is obtained, under the boundary conditions
\begin{eqnarray}
- \left ( \frac{AC p_2 - (BC + O_{12})p_1}{A O_{12}} \right )
(\tau_{\alpha}) \,\, \pi_{i}' (\tau_{\alpha}) + \left ( \frac{A
p_2 - B p_1}{O_{12}} \right ) (\tau_{\alpha}) \,\, {v'}^{i}
(\tau_{\alpha}) & = & V^i_{\alpha} \, , \quad \alpha=1,2; \quad
i=1,2. \nonumber\\
\end{eqnarray}
In this variational principle the phase space variables are $({\vec v}' ,
{\vec \pi}')$ and $u^i (\tau_{\alpha})=U^i_{\alpha}$ are parameters [cf Eq.
(\ref{OAHHI}) in the case of the harmonic oscillator]. Note that in the
boundary term of last action as well as in last boundary conditions $p_i
(\tau_{\alpha})$ is a function of $A$, $B$, $C$, $u^i (\tau_{\alpha})$,
${v'}^{i} (\tau_{\alpha})$, and ${p'}_2 (\tau_{\alpha})$. Therefore the
boundary conditions and the action are well defined. To arrive at this result
the first set of equations in Eq. (\ref{NIBCmod}) was used, which gives $p_i
(\tau_{\alpha})$ as a function of $A$, $B$, $C$, and ${p'}_2 (\tau_{\alpha})$.
Nevertheless, using the constraint $D'=0$, ${p'}_2 (\tau_{\alpha})$ can be put
as a function of $A$, $B$, $C$, ${v'}^i (\tau_{\alpha})$, and ${\pi'}_i
(\tau_{\alpha})$. The final result comes from the combination of these two
partial results. Notice also that the number of boundary conditions has
decreased. This form of the variational principle is very interesting. It
contains a kinematical term like any action, but it has not a Hamiltonian,
that is to say, all its dynamical information has been mapped to its time
boundary. Therefore it is natural to interpret this result as the ``canonical
version" of the holographic hypothesis in the sense that with this particular
choice of the gauge condition, the dynamics of the system is now at the time
boundary \cite{Susskind1995}.

Alternatively, the second option is to solve the equations of
motion which are the original ones but with ${\vec u}$, ${\vec
v}$, ${\vec p}$, ${\vec \pi}$, $N$, $M$, and $\lambda$ replaced by
${\vec u}'$, ${\vec v}'$, ${\vec p}'$, ${\vec \pi}'$, $N'$, $M'$,
and $\lambda'$. Their solution, using the gauge ${u'}^1=A$,
${u'}^2=B$, ${p'}_1 = C$ is
\begin{eqnarray}
{u'}^1 & = & A \, , \quad {u'}^2 = B \, ,\quad {p'}_1 = C \, ,
\quad {p'}^2 = D \, , \nonumber\\
{v'}^1 & = & E \, , \quad {v'}^2 = F \, , \quad {\pi'}_1 = G \, ,
\quad  {\pi'}_2 = H \, . \label{letters}
\end{eqnarray}
with $A \cdots H$ constants ($\tau$ independent), and therefore they are Dirac
observables (of course they are not independent, rather, they are related by
means of the constraint equations). Inserting this solution in the action
(\ref{NIAmod}), the term $S [ {\vec u}' , {\vec v}' , {\vec p}' , {\vec \pi}',
N' , M' , \lambda' ]$ vanishes,  and the only contribution is given by the
boundary term there.

It is worth emphasizing that $D \cdots H$ are indeed Dirac observables. To
obtain the explicit expressions of these observables in terms of the original
phase space variables ${\vec u}$, ${\vec v}$, ${\vec p}$, and ${\vec \pi}$, it
is necessary to use Eqs. (\ref{Variables}) and (\ref{EGP}). From them,
\begin{eqnarray}
{p'}_2 & = & \frac{O_{12} + A C}{A}\, ,\nonumber\\
{\pi'}_1 & = & \frac{- A O_{23} + B O_{13}}{O_{12}}\, ,\nonumber\\
{\pi'}_2 & = & \frac{- A O_{24} + B O_{14}}{O_{12}}\, ,\nonumber\\
{v'}^1 & = & \frac{-A C O_{23} + O_{13} (O_{12} +BC)}{A O_{12}}\, ,\nonumber\\
{v'}^2 & = & \frac{- A C O_{24} + O_{14} (O_{12} +BC)}{A O_{12}}
\end{eqnarray}
(see Ref. \cite{Mon99} for the definition of the $O_{ij}$ observables). But,
because of Eq. (\ref{letters}), these two expressions for ${\vec u}'$, ${\vec
v}'$, ${\vec p}'$, and ${\vec \pi}'$ must be the same thing. Therefore $D$,
$E$, $F$, $G$, and $H$ are given by
\begin{eqnarray}
D & = & \frac{O_{12} + A C}{A}\, ,\nonumber\\
E & = & \frac{- A O_{23} + B O_{13}}{O_{12}}\, ,\nonumber\\
F & = & \frac{- A O_{24} + B O_{14}}{O_{12}}\, ,\nonumber\\
G & = & \frac{-A C O_{23} + O_{13} (O_{12} +BC)}{A O_{12}}\, ,\nonumber\\
H & = & \frac{- A C O_{24} + O_{14} (O_{12} +BC)}{A O_{12}}.
\end{eqnarray}
Of course, these five observables are non independent, there are restrictions
among them \cite{Mon99}. The important point is that this shows that the
dynamics is frozen in this particular gauge.

\subsubsection{Fully gauge-invariant $SL(2,R)$ model}
The simplest boundary term needed to build $S_{\mbox{\small inv}}$ is
$F=\frac{1}{2} \left ( \vec{u} \cdot \vec{p} + \vec{v} \cdot \vec{\pi}\right
)$. Of course an arbitrary function of the Dirac observables $F_1(\phi,
J,\epsilon,\epsilon')$ can been added too. However, particular choices for
$F_1$ just modify the boundary conditions. Therefore the simplest variational
principle has the gauge-invariant action
\begin{eqnarray}
S_{\mbox{\small inv}} [ {\vec u} , {\vec v} , {\vec p} , {\vec \pi}, N , M,
\lambda ] & = & S - \frac{1}{2} \left ( \vec{u} \cdot \vec{p} + \vec{v} \cdot
\vec{\pi} \right ) \mid^{\tau_2}_{\tau_1} \, . \label{inv}
\end{eqnarray}
Due to the fact that the action $S$ has been replaced by $S_{\mbox{\small
inv}}$, the boundary conditions must be modified too. Using Eq.
(\ref{canonical}), the canonical transformation induced by the boundary term
$B_2$ is
\begin{eqnarray}
{\cal Q}^1 & = & \frac{1}{2} \ln{\left ( \frac{u^1}{p_1}\right )}
\, , \quad {\cal P}_1 = u^1 p_1 \, ,\nonumber\\
{\cal Q}^2 & = & \frac{1}{2} \ln{\left ( \frac{u^2}{p_2}\right )}
\, , \quad {\cal P}_2 = u^2 p_2 \, ,\nonumber\\
{\cal Q}^3 & = & \frac{1}{2} \ln{\left ( \frac{v^1}{\pi_1}\right )}
\, , \quad {\cal P}_3 = v^1 \pi_1 \, ,\nonumber\\
{\cal Q}^4 & = & \frac{1}{2} \ln{\left ( \frac{v^2}{\pi_2}\right )} \, ,
\quad {\cal P}_4 = v^2 \pi_2 \, . \label{BBB}
\end{eqnarray}
Thus the new boundary conditions associated with $S_{\mbox{\small inv}}$ are
\begin{eqnarray}
\frac{1}{2} \ln{\left ( \frac{u^1}{p_1}\right )} (\tau_{\alpha}) &
= & {\cal Q}^1_{\alpha} \, ,\nonumber\\
\frac{1}{2} \ln{\left ( \frac{u^2}{p_2}\right )}(\tau_{\alpha})
& = & {\cal Q}^2_{\alpha}\, , \nonumber\\
\frac{1}{2} \ln{\left ( \frac{v^1}{\pi_1}\right )}(\tau_{\alpha})
& = &
{\cal Q}^3_{\alpha}\, ,\nonumber\\
\frac{1}{2} \ln{\left ( \frac{v^2}{\pi_2}\right )}(\tau_{\alpha})
& = & {\cal Q}^4_{\alpha}\, , \quad \alpha=1,2 . \label{boundaryII}
\end{eqnarray}
In summary, the original variational principle is defined by the
action (\ref{inv}) and by the boundary conditions
(\ref{boundaryII}) if $({\vec u} , {\vec p})$, and $({\vec v} ,
{\vec \pi})$ are used as canonical variables. Suppose the boundary
conditions
\begin{eqnarray}
u^1 & = & A \, ,\quad u^2 = B \, , \quad p_1 = C
\end{eqnarray}
were imposed. It is clear that they would conflict with the boundary
conditions. Applying the method, the action in the new variational principle
is simply Eq. (\ref{inv}) but rewritten in terms of $({\vec u}' , {\vec p}')$,
and $({\vec v}' , {\vec \pi}')$
\begin{eqnarray}
S_{\mbox{\small inv}} [ {\vec u}' , {\vec v}' , {\vec p}' , {\vec \pi}', N' ,
M', \lambda' ] & = & S' - {B_2}' \mid^{\tau_2}_{\tau_1}\, , \nonumber\\
& = & S'  - \frac{1}{2} \left ( \vec{u}' \cdot \vec{p}' + \vec{v}'
\cdot \vec{\pi}'
\right ) \mid^{\tau_2}_{\tau_1} \, , \nonumber\\
\label{invII}
\end{eqnarray}
under the boundary conditions
\begin{eqnarray}
\frac{1}{2} \ln{\left ( \frac{{u'}^1}{{p'}_1}\right )}
(\tau_{\alpha}) & = & \frac{1}{2} \ln{\left (
\frac{\beta(\tau_{\alpha}) + \alpha (\tau_{\alpha}) e^{2
{\cal Q}^1_{\alpha}}}{ \delta (\tau_{\alpha}) + \gamma (\tau_{\alpha})
e^{2 {\cal Q}^1_{\alpha}}} \right )}\, ,\nonumber\\
\frac{1}{2} \ln{\left ( \frac{{u'}^2}{{p'}_2}\right
)}(\tau_{\alpha}) & = & \frac{1}{2} \ln{\left (
\frac{\beta(\tau_{\alpha}) + \alpha (\tau_{\alpha}) e^{2
{\cal Q}^2_{\alpha}}}{ \delta (\tau_{\alpha}) + \gamma (\tau_{\alpha})
e^{2 {\cal Q}^2_{\alpha}}} \right )}
\, ,\nonumber\\
\frac{1}{2} \ln{\left ( \frac{{v'}^1}{{\pi'}_1}\right
)}(\tau_{\alpha}) & = & \frac{1}{2} \ln{\left (
\frac{\gamma(\tau_{\alpha}) + \delta (\tau_{\alpha}) e^{2
{\cal Q}^3_{\alpha}}}{ \alpha (\tau_{\alpha}) + \beta (\tau_{\alpha})
e^{2 {\cal Q}^3_{\alpha}}} \right )}
\, ,\nonumber\\
\frac{1}{2} \ln{\left ( \frac{{v'}^2}{{\pi'}_2}\right
)}(\tau_{\alpha}) & = & \frac{1}{2} \ln{\left (
\frac{\gamma(\tau_{\alpha}) + \delta (\tau_{\alpha}) e^{2
{\cal Q}^4_{\alpha}}}{ \alpha (\tau_{\alpha}) + \beta (\tau_{\alpha})
e^{2 {\cal Q}^4_{\alpha}}} \right )}\, , \quad \alpha=1,2 ,
\label{nboundaryII}
\end{eqnarray}
which were obtained rewriting Eq. (\ref{boundaryII}) in terms of $({\vec u}' ,
{\vec p}')$, $({\vec v}' , {\vec \pi}')$, and the gauge parameters.

In the gauge related variables, the gauge condition is
\begin{eqnarray}
{u'}^1 & = & A \, , \quad {u'}^2 = B\, ,\quad {p'}_1 = C \, .
\label{fixingmod}
\end{eqnarray}
By inserting Eq. (\ref{Variables}) and using $\alpha \delta -\beta \gamma =1$
the expressions of the gauge parameters are computed. Of course, they are the
same as those found in the previous subsection, and given by Eq. (\ref{EGP}).
Notice again that $u^i (\tau_{\alpha})$ and $p_i (\tau_{\alpha})$ are
parameters. The difference with the non-gauge-invariant case is that in the
present case these parameters do not appear in the action, they appear in the
boundary conditions only. Therefore the goal has been reached. The new
variational principle is defined by the action (\ref{invII}) with the boundary
conditions (\ref{nboundaryII}) where the gauge parameters are given by Eq.
(\ref{EGP}). In this variational principle the boundary conditions are
compatible with the gauge conditions.

Now, as before, the gauge-fixed variational principle will be computed. To do
this, the gauge conditions (\ref{fixingmod}) and the constraint equations must
be inserted into the action and into the boundary conditions. This gives a
reduced action with, of course, a lower number of boundary conditions. By
doing this, the variational principle is defined by the action
\begin{eqnarray}
S_{\mbox{\small fixed}} [{\vec v}' , {\vec \pi}'] & = & \int^{\tau_2}_{\tau_1}
d \tau \left [ \frac{d {v'}^1}{d \tau} {\pi'}_1 + \frac{d {v'}^2}{d\tau}
{\pi'}_2 \right ] - {\vec v}' \cdot {\vec \pi}' \mid^{\tau_2}_{\tau_1}
\end{eqnarray}
under the boundary conditions
\begin{eqnarray}
\frac{1}{2} \ln{\left ( \frac{{v'}^1}{{\pi'}_1}\right
)}(\tau_{\alpha}) & = & \frac{1}{2} \ln{\left (
\frac{\gamma(\tau_{\alpha}) + \delta (\tau_{\alpha}) e^{2
{\cal Q}^3_{\alpha}}}{ \alpha (\tau_{\alpha}) + \beta (\tau_{\alpha})
e^{2 {\cal Q}^3_{\alpha}}} \right )}
\, ,\nonumber\\
\frac{1}{2} \ln{\left ( \frac{{v'}^2}{{\pi'}_2}\right
)}(\tau_{\alpha}) & = & \frac{1}{2} \ln{\left (
\frac{\gamma(\tau_{\alpha}) + \delta (\tau_{\alpha}) e^{2
{\cal Q}^4_{\alpha}}}{ \alpha (\tau_{\alpha}) + \beta (\tau_{\alpha})
e^{2 {\cal Q}^4_{\alpha}}} \right )}\, , \quad \alpha=1,2 .
\end{eqnarray}
The constraint $D'=0$ was used to reduce the boundary term in the action. In
this variational principle the phase space variables are $({\vec v}' , {\vec
\pi}')$. Notice that there are no parameters in the action, rather, the
parameters $u^i (\tau_{\alpha})$, and $p_i (\tau_{\alpha})$ are in the
boundary conditions. The new thing here is that these parameters can be, using
the first four equations in Eq. (\ref{boundaryII}), the first four equations
in Eq. (\ref{nboundaryII}), and the constraint $D'=0$, put in terms of $A$,
$B$, $C$, ${\cal Q}^1_{\alpha}$, ${\cal Q}^2_{\alpha}$, ${v'}^i
(\tau_{\alpha})$, and ${\pi'}_i (\tau_{\alpha})$. Therefore the variational
principle is well defined, its action has not a Hamiltonian and its dynamics
sits both at the boundaries and in the boundary conditions. This might be
interpreted as the canonical version of the holographic hypothesis
\cite{Susskind1995}.

\section{What is a Hamiltonian constraint?}
Hamiltonian constraints are quadratic in the momenta. This fact implies that a
boundary term arises when the gauge transformed action is computed. On the
other hand, in Sec. II fully gauge-invariant actions were built in spite of
the fact the systems have first class constraints quadratic in the momenta. Is
there anything deep in fully gauge-invariant actions besides their aesthetic
property? Is the gauge symmetry of generally covariant theories with first
class constraints quadratic in the momenta of a different kind with respect to
the one of Yang-Mills theories which have constraints linear in the momenta?
In this section, new variational principles with first class constraints
linear in the momenta will be built for the generally covariant systems with
first class constraints quadratic in the momenta studied in Sec. II. These
variational principles will be written in terms of the new phase space
variables $({\cal Q}^i , {\cal P}_i)$ naturally induced by the boundary term.
According to Eq. (\ref{canonical}) with $B$ replaced by $B+F$ the new
variables are not unique and there is a freedom to select an appropriated
combination of $B+F$ in such way that in the infinitesimal case the system of
Eqs.(\ref{DIFFEQ}) have a solution. Once, a solution of Eqs. (\ref{canonical})
and (\ref{DIFFEQ}) is found the fully gauge-invariant action is given by
\begin{eqnarray}
S_{\mbox{\small inv}}[{\cal Q}^i,{\cal P}_i,\lambda^a] & = &
\int_{\tau_1}^{\tau_2}d\tau \left ( p_i {\dot q}^i-H_E - \frac{d}{d\tau}(B+F)
\right ) \, ,
\end{eqnarray}
subject to the boundary conditions
\begin{equation}
{\cal Q}^i(\tau_\alpha) = {\cal Q}^i_\alpha, \qquad \alpha=1,2.
\end{equation}

\subsection{Parametrized relativistic free particle}
The fully gauge-invariant action associated with the parametrized relativistic
free particle is given by
\begin{eqnarray}
S_{\mbox{\small inv}}[ x^{\mu} , p_{\mu}, \lambda ] & = & \int^{s_2}_{s_1} d s
\left [ \frac{d x^{\mu}}{d s} p_{\mu} - \lambda \left ( p^{\mu} p_{\mu} + m^2
\right ) \right ] - x^{\mu} p_{\mu} \mid^{s_2}_{s_1}\, ,
\end{eqnarray}
under the boundary conditions (\ref{XXX}). The boundary term induces the
canonical transformation (\ref{VFP}) from the original set of variables
$(x^{\mu}, p_{\mu})$ to the new phase space variables $({\cal Q}^{\mu}, {\cal
P}_{\mu})$. By introducing the boundary term $- x^{\mu} p_{\mu}
\mid^{s_2}_{s_1}$ into the integral $S_{\mbox{\small inv}}$ can be written in
terms of the new phase space variables
\begin{eqnarray}\label{parnue}
S_{\mbox{\small inv}} [{\cal Q}^{\mu}, {\cal P}_{\mu} , \lambda ] & = &
\int^{s_2}_{s_1} d s \left [ \frac{d {\cal Q}^{\mu}}{d s} {\cal P}_{\mu} -
\lambda \left ( \beta^2 {\cal Q}^{\mu} {\cal Q}_{\mu} + m^2 c^2 \right )
\right ]
\end{eqnarray}
under the boundary conditions
\begin{eqnarray}
{\cal Q}^{\mu} (s_{\alpha}) & = & {\cal Q}^{\mu}_{\alpha}\, .
\end{eqnarray}
Notice three things: (i) the action written in terms of ${\cal Q}^{\mu}$,
${\cal P}_{\mu}$ has no boundary term, (ii) the first class constraint does
not depend on the momenta ${\cal P}_{\mu}$, (iii) the action (\ref{parnue})
transforms under gauge transformations in similar way to the action for
self-dual gravity \cite{MonVer00}, whereas the transformation properties of
the action (\ref{FPaction}) are equivalent to those of gravity in ADM
variables.

\subsection{Parametrized harmonic oscillator}
The fully gauge-invariant action associated with the parametrized harmonic
oscillator is given by
\begin{eqnarray}
S_{\mbox{\small inv}} [x, t, p, p_t , \lambda ] & = & \int^{\tau_2}_{\tau_1} d
\tau \left [ \frac{d x}{d \tau} p + \frac{d t}{d \tau} p_t - \lambda \left
(p_t + \frac{p^2}{2m} + \frac12 m \omega^2 x^2
\right ) \right ] - \frac{1}{2} x p \mid^{\tau_2}_{\tau_1}  \nonumber\\
\end{eqnarray}
under the boundary conditions (\ref{BC}). The boundary term
induces the canonical transformation (\ref{WWW}) from the initial
canonical variables $(x,t;p,p_t)$ to the new canonical set
$({\cal X,T;P,P_T})$. By introducing the boundary term $- \frac{1}{2} x p
\mid^{\tau=\tau_2}_{\tau=\tau_1}$ into the integral the
gauge-invariant action is written in terms of the new canonical
variables
\begin{eqnarray}
S_{\mbox{\small inv}} [{\cal X ,T , P, P_T} , \lambda ] & = &
\int^{\tau_2}_{\tau_1} d \tau \left [ \frac{d {\cal X}}{d \tau}{\cal P} +
\frac{d {\cal T}}{d \tau} {\cal P_T} - \lambda \left ( {\cal P_T} +
\frac{1}{2m} {\cal P} e^{- 2 {\cal X}} + \frac{1}{2}m
\omega^2 {\cal P} e^{2 {\cal X}} \right ) \right ]\, , \nonumber\\
& & \label{AONEW}
\end{eqnarray}
under the boundary conditions
\begin{eqnarray}
{\cal X}(\tau_{\alpha}) & = & {\cal X}_{\alpha}\, ,\quad {\cal T}(\tau_{\alpha}) =
{\cal T}_{\alpha} \, , \quad \alpha=1,2\, .
\end{eqnarray}
Notice two things: (i) Eq. (\ref{AONEW}) has no boundary term, (ii) the first
class constraint in Eq. (\ref{AONEW}) is linear and homogeneous in the new
momenta ${\cal P}$, ${\cal P_T}$. These two facts are related. Due to the fact
that the action (\ref{AONEW}) is fully gauge-invariant and has no boundary
term, according to Ref. \cite{Henneaux1992} the first class constraint has to
be linear and homogeneous in the momenta as it really happens.

\subsection{$SL(2,R)$ model}
The fully gauge-invariant action which captures the $SL(2,R)$
gauge symmetry of this model is
\begin{eqnarray}
S_{\mbox{\small inv}} [ {\vec u} , {\vec v} , {\vec p} , {\vec \pi}, N , M,
\lambda ] & = & \int_{\tau_1}^{\tau_2} d\tau \left [ \dot {\vec u} \cdot {\vec
p} + \dot {\vec v} \cdot {\vec \pi} - \left ( N H_1 +
M H_2 +\lambda D \right ) \right ] \nonumber\\
& & - \frac12 ( {\vec u} \cdot {\vec p} + {\vec v }\cdot {\vec
\pi} )\mid^{\tau_2}_{\tau_1} \, , \label{SL(2,R)A}
\end{eqnarray}
with the boundary conditions (\ref{boundaryII}). The boundary term induces a
canonical transformation given by Eq. (\ref{BBB}) from the initial set of
canonical variables $({\vec u}, {\vec v}; {\vec p}, {\vec \pi} )$ to the new
one $({\cal Q}^i , {\cal P}_i)$. Then it is possible to rewrite the
variational principle in terms of these new variables. This is done by
introducing the boundary term into the integral in Eq. (\ref{SL(2,R)A}), and
$S_{\mbox{\small inv}}$ becomes
\begin{eqnarray}
S_{\mbox{\small inv}} [{\cal Q}^i , {\cal P}_i , N, M , \lambda ] & = &
\int^{\tau_2}_{\tau_1} d \tau \left [\frac{d {\cal Q}^i}{d \tau} {\cal P}_i -
\left ( N C_1 + M C_2 + \lambda C_3 \right ) \right ] \, , \label{SL(2,R)NA}
\end{eqnarray}
with
\begin{eqnarray}
C_1 & = & \frac12 \left [ {\cal P}_1 e^{-2 {\cal Q}^1}+ {\cal P}_2 e^{-2 {\cal Q}^2} -
{\cal P}_3 e^{2 {\cal Q}^3} - {\cal P}_4 e^{2 {\cal Q}^4} \right ] \, , \nonumber\\
C_2 & = & \frac12 \left [ {\cal P}_3 e^{-2 {\cal Q}^3}+ {\cal P}_4 e^{-2 {\cal Q}^4} -
{\cal P}_1 e^{2 {\cal Q}^1} - {\cal P}_2 e^{2 {\cal Q}^2} \right ]\, , \nonumber\\
C_3 & = & {\cal P}_1 + {\cal P}_2 - {\cal P}_3 - {\cal P}_4 \, .
\end{eqnarray}
under the boundary conditions
\begin{eqnarray}
{\cal Q}^i (\tau_{\alpha}) & = & {\cal Q}^i_{\alpha}\, , \quad i=1,2,3,4\, ,
\quad \alpha=1,2\, .
\end{eqnarray}
Again, the new variational principle in terms of the phase space variables
$({\cal Q}^i , {\cal P}_i)$ features (i) no boundary term in the action
(\ref{SL(2,R)NA}), and (ii) the first class constraints are linear and
homogeneous in the momenta ${\cal P}_i$.

It could be said in some sense that the parametrized harmonic oscillator and
the $SL(2,R)$ model are of the same kind, both have first class constraints
quadratic in the configuration and momentum variables. In both cases, the
simplest boundary term needed to build $S_{\mbox{\small inv}}$ is $-\frac12
q^i p_i$. The parametrized relativistic free particle, on the other hand, is
quadratic in the momenta only. In that case, the boundary term needed to build
$S_{\mbox{\small inv}}$ is $- x^{\mu} p_{\mu}$ and $S_{\mbox{\small inv}}$ is
fully gauge-invariant on the constraint surface only. In all cases, when
$S_{\mbox{\small inv}}$ is written in terms of the phase space variables
induced by the boundary term it happens that no boundary term is present
anymore. In the cases where $S_{\mbox{\small inv}}$ is fully gauge-invariant
[parametrized harmonic oscillator and the $SL(2,R)$ model] the new constraints
are {\it linear} and {\it homogeneous} in the new momentum variables whereas
in the parametrized relativistic free particle where $S_{\mbox{\small inv}}$
is gauge-invariant only on the constraint surface the new constraint is {\it
not} homogeneous in the new momenta.

\section{Concluding remarks}
In this paper two main topics were touched on. The first one was the
implementation of the ideas of Ref. \cite{Henneaux1992} for gauge systems when
{\it finite} gauge transformations are taken into account to make compatible
both boundary and gauge conditions. In this case, also two different
variational principles were analyzed. The first one features
non-gauge-invariant actions whereas the other includes fully gauge-invariant
ones. One of the main lessons learned from the finite but non-gauge-invariant
approach is that the interplay between constraints quadratic and linear in the
momenta can result in a contribution of the seconds to the boundary term in
contrast to the infinitesimal approach. The second contribution was to take
advantage of fully gauge-invariant actions to rewrite such systems in terms of
new phase space variables in terms of which the first class constraints are
linear (and homogeneous in some cases) in the momenta. For a long time it has
been considered that first class constraints quadratic in the momenta are the
distinguishing mark of generally covariant theories, as general relativity or
string theory [see, for instance, Ref. \cite{Teitelboim1982}]. Here it was
shown that these kind of theories can be written in terms of first class
constraints linear in the momenta if fully gauge-invariant actions are taken
into account. Thus the gauge symmetry present in generally covariant theories
with first class constraints quadratic in the momenta is apparently of the
same kind as the gauge symmetry present in Yang-Mills theories if fully
gauge-invariant actions are taken into account for the former: after all, both
kinds of theories can be described with first class constraints linear in the
momenta. It is important to recall that some steps in this direction have been
done for Bosonic strings, at least infinitesimally \cite{MonReport}. In the
case of general relativity, it has been shown in Ref. \cite{MonVer00} that the
action for gravity in terms of Ashtekar variables
\cite{Pleb77,Sam87,Jacob88,Capo91a,Abhay} is gauge-invariant up to a term
proportional to the Hamiltonian constraint under the gauge symmetry generated
by their first class constraints. This suggests that if one starts from the
action for gravity in terms of Arnowitt-Deser-Misner (ADM) variables in its
triad version \cite{Lect} and apply Eq. (\ref{DIFFEQ}), the solution for the
generating function will be the one introduced in Ref. \cite{Dolan}. From the
models studied here it is concluded that gauge invariance for the actions is
the reason to have first class constraints linear in the momenta.

There are other aspects of this last topic which were not analyzed here but
deserve to be studied. At a classical level, the introduction of the boundary
term into the integral actions could have also been handled with the original
set of canonical variables instead of making the canonical transformation as
it was done here, but this other way of dealing with the boundary term in the
fully gauge-invariant actions would have led to the introduction of second
class constraints. On the other hand, it would be worth analyzing the quantum
theory emerging from these systems with first class constraints linear in the
momenta (and homogeneous in some cases) and compare with the standard
quantization coming from their quadratic constraints. These issues are left
for future work.

\section*{Acknowledgments}
M.M. thanks financial support provided by the {\it Sistema Nacional de
Investigadores} of the Secretar\'{\i}a de Educaci\'on P\'ublica (SEP) of
Mexico. M.M. was also supported by grant CINVESTAV JIRA'2001/16. J.D.V. was
partially supported by grants DGAPA-UNAM IN100397 and CONACyT 32431-E.


\end{document}